\documentstyle[prl,aps,epsf]{revtex}
\begin{document}
\draft

\twocolumn[\hsize\textwidth\columnwidth\hsize\csname @twocolumnfalse\endcsname

\title{Melting of Moving Vortex Lattices in Systems With Periodic Pinning} 
\author{C.~Reichhardt and G.T.~Zim\'anyi} 
\address{Department of Physics, University of California, Davis,
Davis, CA 95616}

\date{\today}
\maketitle
\begin{abstract}
We study numerically the effects of temperature 
on moving vortex lattices interacting with  
periodic pinning arrays. For low temperatures the vortex lattice 
flows in channels, forming a hexatic structure with long range  
transverse and longitudinal ordering.  
At higher temperatures, a transition to a smectic state occurs
where vortices wander between channels 
and longitudinal order is lost while transverse order remains. 
At the highest temperatures the vortex lattice melts into an isotropic
liquid. 
\end{abstract}
\vspace{-0.1in}
\vspace{-0.2in}
\vskip2pc]

Driven vortex lattices interacting with  
quenched and thermal disorder are an ideal system in which to study
the nonequilibrium phases and transitions that arise from the interplay of 
competing interactions 
\cite{Koshelev,Giamarchi,Brag-Balents-Scheidl,Olson,Higgins-Beasly,Yaron,Kes}. 
Transport measurements \cite{Higgins-Beasly}, neutron scattering \cite{Yaron}, 
and Bitter decoration experiments \cite{Kes} 
have provided strong evidence for transitions between different 
vortex dynamic phases, including creep, plastic flow, and  
ordered (elastic) flow. Theoretical work and simulations \cite{Koshelev}
suggested that for a {\it moving} vortex lattice,
at high velocities the effect of disorder can be represented via a shaking
temperature $T_{sh}$, inversely proportional to the velocity. 
At high velocities $T_{sh}$ decreases below the melting temperature
$T_{m}$, and the vortices reorder.  $T_m$ 
of the ordered moving vortex lattice is near the equilibrium 
$T_m$ of the disorder-free stationary lattice.   
In the highly driven, solidified state, theoretical \cite{Giamarchi}
work suggested that the vortex lattice forms a moving Bragg-glass (MBG),
or a strongly anisotropic moving smectic (MS) or 
moving transverse glass (MTG) \cite{Brag-Balents-Scheidl}. 
In the MBG the vortices move in correlated channels with few defects,
producing quasi-long range order. 
In the MS/MTG vortices move in uncorrelated channels, so although  
power law transverse order is present, the longitudinal order 
is short range only. Strongly anisotropic ordering, consistent 
with a MS/MTG, as well as more ordered phases, consistent with a MBG and 
vortex channeling, have been seen in simulations \cite{Olson} 
and Bitter decoration experiments \cite{Kes}.  

Despite the considerable work done in the case of random quenched disorder, 
the effects of thermal disorder and melting for the interesting case of 
a moving vortex lattice interacting with a 
{\it periodic} pinning substrate have been far less studied.  
Periodic pinning substrates in superconductors can be created with
arrays of microholes \cite{Baert-holes} and magnetic
dots \cite{Schuller}. 
In all such systems,  
the pin radius is 
smaller than the distance between pins, so that a moving 
vortex spends the largest fraction of its time in the unpinned area. 
A crucial difference from the random pinning case
is that the effect of the periodic pinning cannot
be represented by a shaking temperature.  The same periodicity
also induces true long range transverse or longitudinal order in the 
moving lattice. 

We report a numerical study of the melting of moving 
vortices in square periodic pinning arrays. 
With no driving, the system with pinning 
melts at a {\it higher} temperature
than the pin free system.   
For moving vortices at low temperatures, we observe a   
triangular lattice flowing in strict 1D correlated channels,
with transverse and longitudinal long range order.  The 
transverse order is greater than the longitudinal order, 
and the anisotropy increases with temperature. At higher temperatures 
near the melting temperature of the {\it clean} equilibrium system,
we observe a transition to a moving smectic (MS) state, where 
transverse vortex wandering between channels occurs. 
At even higher temperatures near the melting
temperature of the {\it pinned} equilibrium system,
the moving smectic melts into a moving isotropic liquid (ML). 
We present the dynamic phase diagram and explain its features in terms
of the pinned and unpinned equilibium melting transitions. 

We use finite temperature 
overdamped molecular dynamics simulations in two dimensions. 
${\bf f}_{i} = {\eta}{\bf v}_{i} = {\bf f}_{i}^{vv} + 
{\bf f}_{i}^{vp} + {\bf f}_{d} 
+ {\bf f}_{i}^{T}$. We impose periodic boundary condition in $x$ and
$y$. 
The force between vortices at ${\bf r}_{i}$ and ${\bf r}_{j}$ is  
${\bf f}_{i}^{vv} = \sum^{N_{v}}_{j=1} 
f_{0}A_{v}K_{1} (|{\bf r}_{i} - {\bf r}_{j}|/\lambda){\bf {\hat r}}_{ij}$, 
where $K_{1}(r/\lambda)$ is 
a modified Bessel function, 
$f_{0} = \Phi_{0}^{2}/8\pi^{2}\lambda^{3}$, $\lambda$ is the penetration 
depth and the parameter 
$A_{v}$ tunes the vortex-vortex interaction strength. 
For most of the results here $A_{v} = 1.0$. 
The driving force ${\bf f}_{d}$, 
representing a Lorentz force, is in the $x$ direction. 
The pinning is modeled as attractive parabolic traps of 
maximum strength $f_{p}$ and a range 
$r_{p}$ which is less than the distance between pins $a$.    
The thermal force $f_{i}^{T}$ has the properties $<f_{i}^{T}> = 0$ and
$<f_{i}^{T}(t)f_{j}^{T}(t^{'})> = 2\eta k_{B}T\delta_{ij}\delta(t - t^{'})$. 
The temperature 
$T = (1/2\eta k_{B})(Af_{0})^{2}\delta t $ where $\delta t$ is the time step
in the simulation and $A$ is the number we tune to vary $T$, with 
$f_{i}^{T}=Af_0$. 
We take $f_{0} = k_{B} = \eta = 1$ and use $\delta t = 0.04$. 
We explore the phase diagram by conducting constant $f_{d}$ or $T$ sweeps on
the $f_{d}-T$ plane.
Our model is most relevant to superconductors with 
perodic arrays of columnar defects or thin-film superdonctors where
the vortices can be approximated as 2D objects. A realistic 3D model
would be needed to address the exact nature of the liquid phase,
such as whether it is a line liquid. 
In 

\begin{figure}
\centerline{
\epsfxsize=3.5in
\epsfbox{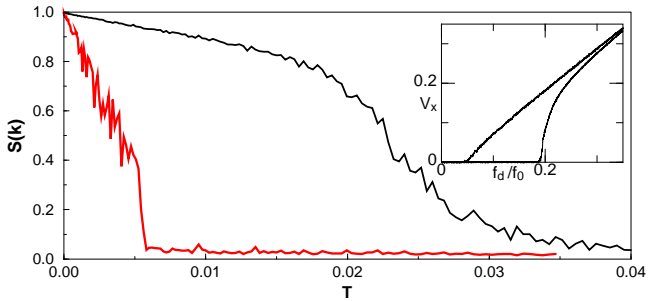}}
\caption{
The structure factor peaks $<S(k)>/N_{v}$ versus
$T$ for a system with $N_{v} = N_{p}$. Upper curve: Sample with a 
square pinning substrate of strength
$f_{p}/f_{0} = 0.22$. Lower curve: Sample containing no pinning. 
The melting for the pinned system 
occurs at much higher $T$ than melting in the clean system.
Inset:  $V_{x}$ versus $f_{d}$ with 
curves for $f_{p}/f_{0} = 0.22$ and $T = 0.0001$ and $0.0725$.} 
\label{fig1}
\end{figure}

\hspace{-13pt} this work we consider only the commensurate case where
the number of vortices $N_{v}$ equals the number of pinning sites $N_{p}$.
Results 
for the incommensurate cases will be presented elsewhere.
The initial vortex positions are generated by simulated annealing 
with each pinning site capturing one vortex.
For finite size scaling we analyze system sizes with 
$N_{v} \sim L^{2}$ for $N_{v} = 224$ to $N_{v}= 2112$. 

We first establish the melting temperature in the pin free and pinned 
systems without external drive ($f_{d} = 0$) 
with
the vortex displacements
$d_{r} = <|({\bf r}(t) - {\bf r}(0))|^{2}>$, and the structure factor 
$S({\bf k}) = \frac{1}{L^{2}}\sum_{i,j}e^{i{\bf k}\cdot[{\bf r}_{i}(t) -
{\bf r}_{j}(t)]}. $  In the pinned system the vortex lattice has 
the same square symmetry as the pinning lattice. 
The melting temperature is determined from the
simultaneous onset of diffusion and a drop in the peaks in $S(k)$. 
In Fig.~1 we show that the melting temperature $T_{m}$ 
is {\it higher} in the pinned system, with 
$T_{m}^{p} \approx 0.03$, than in the unpinned system,
$T_{m}^{np} \approx 0.0058$.
This is reasonable  
since at commensuration the pins stiffen the vortex lattice. 

Next we explore the dynamic phases of the system. For 
$f_{p} = 0.22f_{0}$ the $T = 0$ depinning occurs at $f_{d} = 0.22f_{0}$.
For fixed  $f_{d} = 0.45$ we perform a $T$ sweep, and 
monitor $S({\bf k})$. Every 400 MD steps we measure 
the transverse displacements 
$d_{y} = <|y(T) - y(0)|^{2}>$ from the initial positions of the vortices
at $T = 0$. 
For low drives the vortices form a pinned {\it square} vortex lattice.
The pinned phase is defined by measuring 
$V_{x} = (1/N_{v})\sum^{N_{v}}_{i=1}{\bf v}_{i}\cdot{\bf {\hat x}}$.
In the inset of Fig.~1 we show the 
typical $V_{x}$ versus $f_{d}$ curves for 
two different temperatures.
The transition from the pinned to moving phases are marked by  
a jump in the $V_{x}$ at a well defined $f_{d}$.  
From $V_{x}$ versus $f_{d}$ curves we found little evidence for
plastic or collective creep behaviour 
for temperatures below $T_{m}^{p}$; however, much longer time scales would
be necessary to 
explore
the creep behaviour. 

Above some $T$ dependent driving force, we find that the vortices form 
a {\it triangular} lattice, with the principle lattice vector aligned with
the direction of motion, 
since the vortices spend part of thier time between 
pins where 

\begin{figure}
\centerline{
\epsfxsize=3.5in
\epsfbox{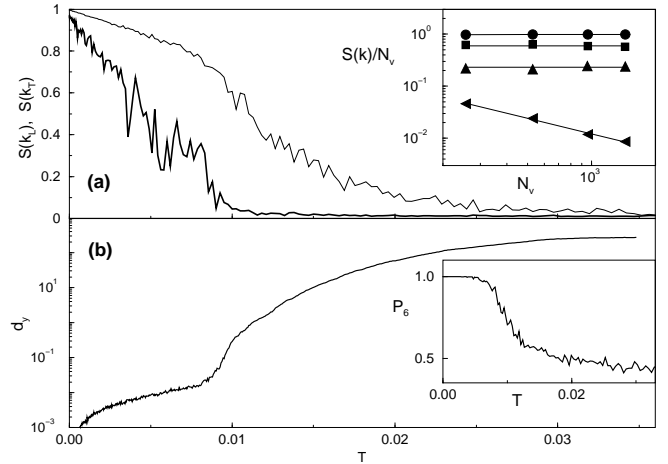}}
\caption{ 
(a) The transverse structure factor peak $S(k_{T})/N_{v}$ (upper curve) and 
the longitudinal peak $S(k_{L})/N_{v}$ (lower curve) for 
a $48\lambda \times 48\lambda$ system with $f_{d} = 0.45f_{0}$, 
$f_{p} = 0.22f_{p}$, and $N_{v} = N_{p}$. (b) The average 
transverse displacements of the vortices from their initial 
positions, $d_{y} = <|y(T) - y(0)|^{2}>$. Inset of (a): the scaling
of $S(k)/N_{v}$ for different system sizes where $N_{v} \sim L^{2}$. 
For T = 0.0025 both the transverse (circles) and longitudinal peaks 
(squares and shifted down by 0.3 for clarity) scale as 
$S(k) \sim L^{0}$; for T = 0.015 the transverse peaks (up triangles) 
scale as $S(k_{T}) \sim L^{0}$ while the longitudinal peaks 
(left triangles) scale as $S(k_{L}) \sim L^{-1.96}$.   
Inset of (b): the evolution of the fraction of six-fold coordinated
vortices $P_{6}$ versus temperature. 
} 
\label{fig2}
\end{figure}

\hspace{-13pt} the vortex-vortex interaction dominates. 

In Fig.~2(a) we plot the transverse $S(k_{T})$ and 
longitudinal $S(k_{L})$ structure function peaks for increasing
$T$. For low $T$, $S(k_{T})$ is only slightly larger than $S(k_{L})$. This 
anisotropy between the peaks becomes more pronounced as $T$ is increased. 
The ordering can be analyzed from the finite size scaling of 
the structure factor as $S(k)/L^{2} \sim L^{-\eta}$ \cite{Franz}. A solid with 
long range order will have $\eta = 0$ while a system with short range
order such as a liquid will have $\eta = 2$. In the inset of Fig.~2(a),  
$S(k_{T})/L^{2}$ and $S(k_{L})/L^{2}$ (where $N_{v} \sim L^{2}$) 
are plotted for different system sizes in the low temperature 
regime ($T = 0.0025$). Both peaks scale as $\eta \approx 0$,
indicative of long range order.  Near $T = 0.0095$, which we label $T_{MS}$,  
$S(k_{L})$ drops precipitously, indicating complete loss of longitudinal 
order, while $S(k_{T})$ retains a significant finite value. 
Gradually $S(k_{T})$ drops to the level of $S(k_{L})$ near $T = 0.03$, 
which we label $T_{ML}$.  In this $T > T_{MS}$ regime the longitudinal 
peak $S(k_{L})/L^{2}$ scales as $\eta = 1.95 \pm 0.03$, consistent 
with an $\eta = 2$ scaling behavior indicating the loss of longitudinal order. 
At this same temperature the transverse peak $S(k_{T})/L^{2}$ shows a 
$\sim L^{-0.0}$ form (triangles in upper inset of Fig.~2), 
indicating that long range transverse order is still present.
The behavior of the two peaks for $T_{MS} < T < T_{ML}$ indicates the
presence of a moving smectic phase. For $T> T_{ML}$ both 
$S(k_{L})$ and $S(k_{T})$ scale as
$\sim L^{-2}$ as the system becomes an isotropic liquid. 
We label these three phases the moving crystal 

\begin{figure}
\centerline{
\epsfxsize=3.5in
\epsfbox{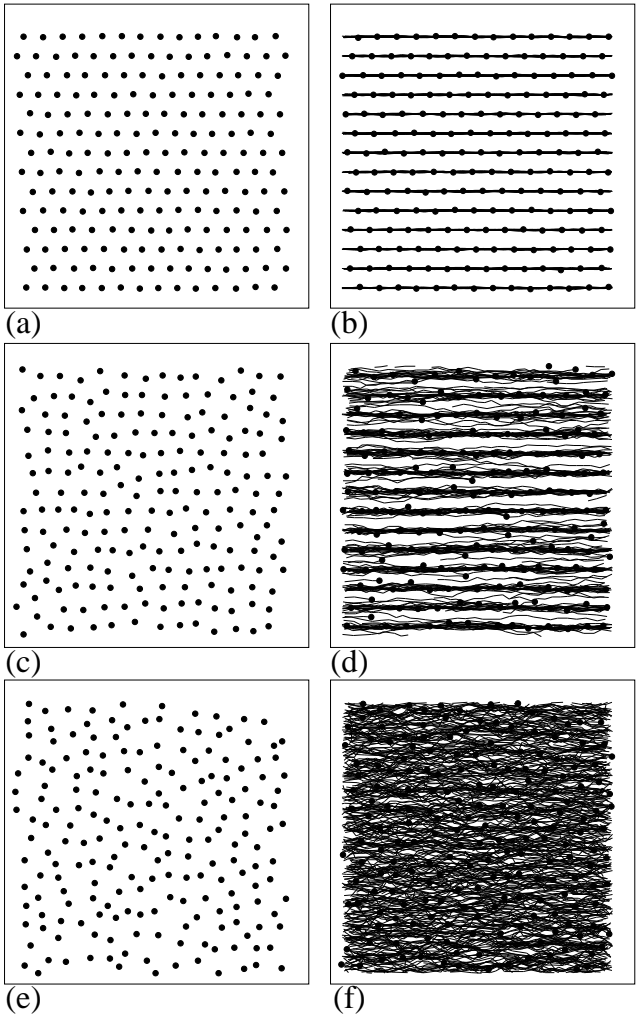}}
\caption{
The real space vortex images (a,c,e) and vortex trajectories (b,d,f) 
for the same system as in Fig.~2 for the moving lattice (T = 0.0025) 
(a,b); the moving smectic (T = 0.015) (c,d); and the moving liquid 
(T = 0.035) (e,f).}  
\label{fig3}
\end{figure}

\hspace{-13pt} `(MC), the
moving smectic (MS) and moving liquid phase (ML). $\eta$  has some 
temperature dependence near the 
transitions which we will
examine elsewhere. In the inset to Fig.~2(b) we plot 
the fraction of 6-fold coordination number $P_{6}$ versus $T$ as 
obtained from the Voronoi construction. The proliferation
of defects occurs at $T = T_{MS}$ as seen by the drop in $P_{6}$.  
In Fig.2(b) we plot the transverse displacement $d_{y}$.  
For $T < T_{MS}, $ $d_{y} \approx 0$ indicating that the 
vortices are moving in straight 1D channels.  
For $T \geq T_{MS}$, $d_{y}$ increases indicating that the onset of 
vortex wondering in the $y$ direction is correlated with the loss of 
longitudinal order and the proliferation of defects. 

To examine
the individual vortex behavior 
in the different dynamic phases 
in Fig.~3 we plot 
snapshots of the vortex positions and trajectories for MC at $T = 0.0025$,
MS at $T = 0.015$, and ML at $T = 0.035$.  In the MC phase, the vortices 
form an ordered triangular lattice [Fig.3(a)], and move  
in correlated 1D paths [Fig.3(b)], which agrees 

\begin{figure}
\centerline{
\epsfxsize=3.5in
\epsfbox{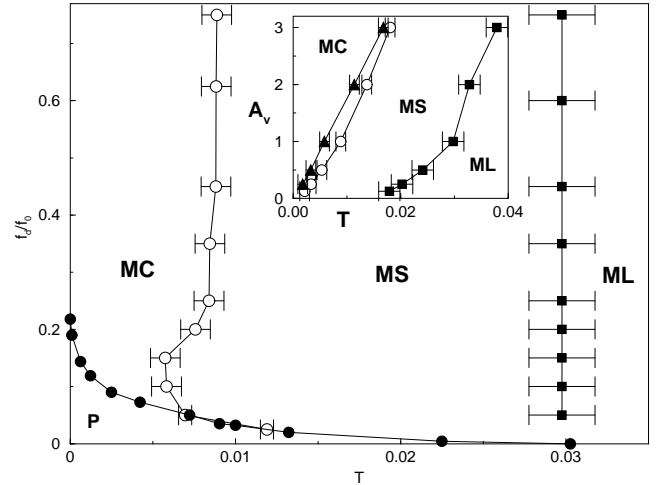}}
\caption{
The dynamic phase diagram of $f_{d}$ versus $T$.
For open circles and 
filled squares, 
the current is fixed 
and the temperature is increased. 
For the filled circles,
the temperature is fixed and the drive is 
increased. The pinned to moving phases are determined from $f_{d}$ versus 
vortex velocity $V_{x}$. The boundaries of the other phases are 
determined from transverse and longitudinal $<S(k)>$ and $d_{y}$. 
The open circles correspond to the MC to ML 
transition and the filled squares correspond to the MS to ML transition for a
system with the same parameters in Fig.~2. The filled circles 
correspond to the transition out of the pinned phase P. 
Inset: The phase diagram for $A_{v}$, the vortex-vortex 
interaction prefactor, versus $T$. The triangles  correspond to the zero drive 
clean melting temperature while the open circles and squares correspond 
to the MC-ML transition and the MS-ML transition with $f_{d} = 0.45f_{0}$.  
}  
\label{fig4}
\end{figure}

\hspace{-13pt} with the zero 
transverse wandering $d_{y} \approx 0$ in Fig.~2(b). 
As $T$ is increased in the MC phase the 
channel width 
increases; 
however, vortices {\it do not cross} from one channel to another.
The pinning also induces the long-range order 
seen from the scaling of $S(k)$, 
whereas algebraic decay would be expected
since the system is in 2D.

In the MS phase, shown in Fig.~3(c), the vortices are much less ordered 
than in the MC phase; however, some order remains.  
The vortex trajectories in the MS phase, Fig.~3(d), reveal that although 
some channeling occurs along the pinning, there is considerable 
vortex motion {\it between} and {\it across} the channels. This 
interchannel vortex motion accounts for the increase of 
the transverse displacements, $d_{y}$, at the onset of the MS phase in
Fig.~2(b). The residual vortex channeling 
accounts for the finite value of $S(k_{T})$ in the MS phase, but the vortex 
positions are uncorrelated between channels so there is
no longitudinal order and the vortex lattice is highly defected. 
Since the residual channels have the same period as the pinning lattice 
the transverse scaling in $S(k)$ gives long range order. 
Finally in the ML phase, shown in Fig.3(e,f), the
vortex positions are disordered and the channeling behavior is lost. 

In Fig.~4 we present the central result of this work, the  
dynamic phase diagram as a function of $f_{d}$ and 
$T$ obtained from fixed drive increasing $T$ simulations (denoted by  
solid circles) and from fixed temperature increasing drive simulations 
(denoted by open circles and squares). The depinning line is determined 
from the pronounced upward curvature in the $V(I)$ curves or the onset 
of displacements. The MC-MS line is found from 
the saturation of $S(k_{L})$ to a minimum value 
as well as the onset of the transverse
displacements. The MS-ML line is obtained from 
the point at which the transverse 
$S(k_{T})$ peak drops to a minimal value and exhibits a 
scaling of $\eta \approx 2$. The MS-ML line is 
roughly independent of drive above $f_{d} = 0.25f_{0}$. The MS line 
shows some curvature toward lower temperatures for drives near the 
depinning line. The MC-MS transition occurs a small amount {\it above} 
$T_{m}^{np} = 0.0058$ but approaches this value for the lower drives. 
The MS-ML transition line coincides with the pinned 
(zero driving value) melting temperature $T_{m}^{p}$. The pinned phase  
vanishes at $T_{m}^{p}$.

The onset of these different phases can be understood by considering that 
for higher drives the moving vortices spend a great part of their time 
{\it outside} the pinning sites because $r_{p} \lesssim a$. 
For temperatures above $T_{m}^{np}$ the vortices enter a molten state
while moving {\it between} pinning sites since they are essentially moving in 
a clean system. In this melted state, the thermal 
fluctuations overcome the vortex-vortex interaction 
and the correlation of vortices in adjacent channels as well
as the longitudinal order is lost. The vortices start diffusing at random,
leading to a large increase in $d_{y}$.
Unlike the case of random pinning, which induces an additional shaking
temperature that effectively {\it lowers} the temperature at which the
vortices disorder or melt, vortices moving in a periodic pinning array 
will {\it not} experience a shaking temperature. 
As long as  $T$ is {\it less} than the melting temperature of the 
non-driven clean system $T_{m}^{np}$, at any drive the overall moving vortex 
lattice remains ordered. Further there is still a pinning effect in 
the transverse direction. This transverse pinning which causes the 
channeling has been seen in simulations with random pinning \cite{Olson} 
and is particularly large for simulations with periodic pinning arrays 
\cite{Reichhardt}.  During the time the vortices are {\it in} the pinning 
sites they still feel a transverse pinning force until $T >T_{m}^{p}$, 
so some vortex channeling persists and finite transverse ordering appears. 
Above $T_{m}^{p}$ the pinning is no longer effective so channeling and 
transverse order are both lost.    

A test of the above interpretation is that $T_{MS}$ is related to 
$T_{m}^{np}$. 
$T_{m}^{np}$
can be tuned by changing
the vortex-vortex interaction prefactor $A_{v}$.  
As $A_{v}$ is lowered $T_{m}^{np}$ will also be lowered.
In the inset of Fig.~4 we show the melting lines for the clean
non-moving system and the
moving system with pinning ($f_{d} = 0.45f_{0}$) for $T$ versus $A_{v}$.  
As $A_{v}$ is increased the $T_{m}^{np}$ and $T_{MS}$ lines increase at 
the same rate. This correlation is encouraging evidence for the above 
interpretation.  For the stiffer lattices (high $A_{v}$) the MC 
phase is larger. 

In summary, we have studied numerically the melting and dynamic phase 
diagram of a vortex lattice interacting with a square periodic pinning 
array. The melting temperature for the non-driven pinned system is 
higher than that for the equivalent clean system.  For moving vortices at 
low $T$ the vortex lattice moves in correlated 1D channels and has 
long range order with some anisotropy between the longitudinal and 
transverse peaks. At a higher temperature there is a transition to a 
Moving Smectic phase where longitudinal order is lost while long-range  
transverse order remains. In the Moving Smectic phase the channel flow 
still persists but vortices diffuse between channels. At high temperatures 
the transverse order and channeling is also lost. 
The Moving Crystal-Moving Smectic transition corresponds roughly to the 
melting temperature of the zero drive clean system. 

We thank C.J. Olson and R.T. Scalettar for 
useful discussions and critical reading of the 
manuscript. Funding provided by CLC/CULAR.

-{\it Note Added} After submission we became aware of the 
paper by V.~Marconi and D.~Dominguez \cite{Marconi} in which they
also study the melting of moving vortex lattices intercting with
a periodic substrate in a peroidc Josephson-juntion array 
in which we find several overlapping results. 
They, however, do {\it not} find a moving smectic phase
which is due to their pinning being modeled as an egg-carton potential
unlike our model in which the space between the pinned sites is 
essential for the smectic phase to occur.







\end{document}